\documentclass[epj]{svjour}
\usepackage{graphicx,latexsym,amssymb,amsmath,color,multirow,mathrsfs,booktabs,ifpdf}
\usepackage{dcolumn}
\usepackage{bm}
\usepackage{float,varioref}
\usepackage{longtable}
\def\aA{$\alpha$-nucleus\ }
\def\AA{nucleus-nucleus\ }
\def\oc{$^{16}$O+$^{12}$C\ }
\def\cc{$^{12}$C+$^{12}$C\ }

\begin{document}
\title{Suppression of the nuclear rainbow in the inelastic \AA scattering}
\author{Nguyen Hoang Phuc\inst{1}\and Dao T. Khoa\inst{1}\and 
Nguyen Tri Toan Phuc\inst{1,2}\and Do Cong Cuong\inst{1}}
\institute{Institute for Nuclear Science and Technology, VINATOM \\ 
179 Hoang Quoc Viet, Cau Giay, 100000 Hanoi, Vietnam \and  
Department of Nuclear Physics, University of Science, VNU-HCM \\
 227 Nguyen Van Cu, District 5, 700000 Ho Chi Minh City, Vietnam}
\date{Received: date / Revised version: date}
\begin{abstract}
{The nuclear rainbow observed in the elastic \aA and light heavy-ion scattering 
is proven to be due to the refraction of the scattering wave by a deep,  
attractive real optical potential. The nuclear rainbow pattern, established as 
a broad oscillation of the Airy minima in the elastic cross section, originates 
from an interference of the refracted far-side scattering amplitudes. It is 
natural to expect a similar rainbow pattern also in the inelastic scattering of a 
\AA system that exhibits a pronounced rainbow pattern in the elastic channel. 
Although some feature of the nuclear rainbow in the inelastic \AA scattering was 
observed in experiment, the measured inelastic cross sections exhibit much 
weaker rainbow pattern, where the Airy oscillation is suppressed and smeared out. 
To investigate this effect, a novel method of the near-far decomposition of the 
inelastic scattering amplitude is proposed to explicitly reveal the coupled 
partial-wave contributions to the inelastic cross section. Using the new decomposition 
method, our coupled channel analysis of the elastic and inelastic \cc and \oc 
scattering at the refractive energies shows unambiguously that the suppression 
of the nuclear rainbow pattern in the inelastic scattering cross section is caused
by a destructive interference of the partial waves of different multipoles. However, 
the inelastic scattering remains strongly refractive in these cases, where the  
far-side scattering is dominant at medium and large angles like that observed 
in the elastic scattering.} 
\end{abstract}
\PACS{       
     {24.10.Ht}{Optical and diffraction models} \and
     {25.55.Ci}{Elastic and inelastic scattering} \and
     {25.70.Bc}{Elastic and quasielastic scattering}   
		} 
\maketitle

\section{Introduction}
\label{sec1} 
The optical model (OM) studies of elastic heavy-ion (HI) scattering usually shows
a strong absorption that suppresses the refractive (large-angle) scattering, and 
elastic HI scattering occurs mainly at the surface. However, with the nuclear 
rainbow pattern observed in the elastic scattering of some $\alpha$-nucleus and 
light HI systems, the absorption turns out to be much weaker and the refractive, 
far-side scattering becomes dominant at medium and large angles \cite{Bra96,Bra97,Kho07r}. 
Not only a fascinating semiclassical analog to the atmospheric rainbow, the nuclear 
rainbow also enables the determination of the real \AA optical potential (OP) with 
much less ambiguity, down to small internuclear distances \cite{Kho07r}. The pattern 
of the nuclear rainbow is usually characterized by a broad oscillation of the Airy 
minima \cite{Bra96,Bra97,Kho07r} in the elastic cross section. The observation of these 
minima, especially, the first Airy minimum A1 that is followed by a broad shoulder-like 
maximum, not only facilitates the determination of the real OP but also provides 
an useful probe of the cluster structure of light nuclei \cite{Mi00-2,Ohk16}. 

Similarly to the atmospheric rainbow, the nuclear rainbow can be interpreted as 
a pattern resulted from an interference of the two scattering amplitudes, as shown
by the barrier-wave/internal-wave (BI) or near-side/far-side (NF) decomposition 
of the elastic scattering amplitude. The BI method proposed by Brink and Takigawa 
\cite{Bri77} describes elastic HI scattering in terms of the internal waves penetrating 
through the potential barrier into the nuclear interior, and barrier waves reflected from 
the barrier. On the other hand, the NF decomposition suggested by Fuller \cite{Ful75} 
splits the elastic scattering amplitude into the near-side (N) and far-side (F) 
components, corresponding to the waves deflected to the near side and far side 
of the scattering center, respectively. These two interpretations are complementary, 
and the broad Airy oscillation of the nuclear rainbow pattern is given by the 
interference of two far-side amplitudes \cite{Kho07r,Mi00,Ann01,Row77}. These are 
either the barrier $f_{\rm BF}$ and internal $f_{\rm IF}$ far-side amplitudes 
\cite{Bri77}, or $f_{{\rm F}_<}$ and $f_{{\rm F}_>}$ far-side amplitudes with the 
orbital momenta $L$ smaller or larger than a critical value $L_{\rm R}$ associated 
with the rainbow angle $\theta_{\rm R}$ \cite{Ful75}.  

From such a pattern of the nuclear rainbow, a similar Airy structure is expected to be 
seen also in the inelastic scattering of a light HI or $\alpha$-nucleus system that 
shows a strong rainbow pattern in the elastic scattering. In fact, some feature of 
the nuclear rainbow was observed in the inelastic light HI scattering 
\cite{Bohlen82,Kho87,Kho05,Mi04,Ohk13,Ohk14,Dem87,Dem92,DAg94,DAg95}, and some of these
data were analyzed using the BI \cite{Mi04} and NF \cite{Dem87,Dem92,DAg94,DAg95} 
decomposition methods. Although these studies have confirmed the dominance of the 
far-side scattering at large angles, the Airy oscillation pattern could not be clearly 
identified in the inelastic cross section. 
\begin{figure}[bht]\vspace*{-0.2cm}\hspace*{-0.2cm}
\includegraphics[width=0.5\textwidth]{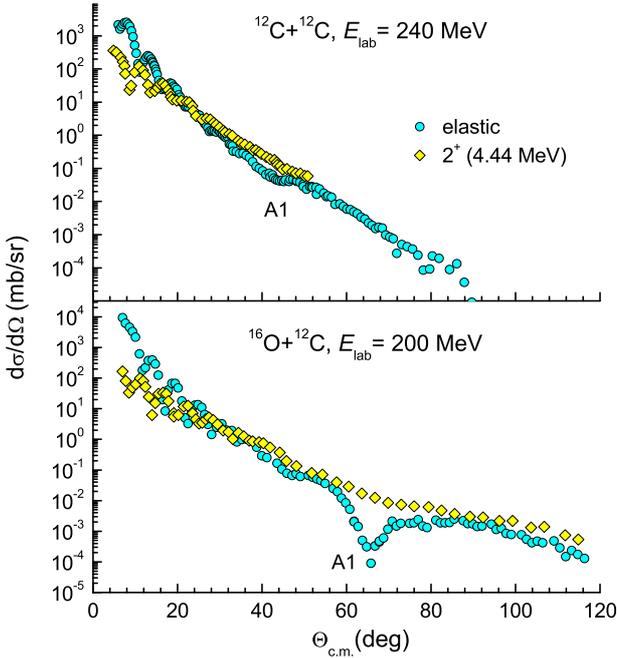}\vspace*{0cm}
\caption{Angular distributions of the elastic scattering and inelastic scattering 
to the $2^+_1$ state of $^{12}$C measured for the \cc system at $E_{\rm lab}=240$ MeV
\cite{Boh85,Dem10} (upper part) and \oc system at $E_{\rm lab}=200$ MeV 
\cite{Ohk14,Ogl00} (lower part). A1 indicates the location of the first Airy
minimum established in the extended OM analysis \cite{Kho16} of the elastic 
scattering data.} \label{f1}
\end{figure}
Such suppression of the nuclear rainbow was assumed by Khoa \emph{et al.} \cite{Kho05} 
to be due to the increased absorption in the exit channel, with the target nucleus 
in an excited state. Dem'yanova \emph{et al.} \cite{Dem92} suggested that 
this effect is linked to the shape of the inelastic form factor, which is different 
from that of the attractive, deep real OP. We note that the inelastic scattering to 
the $0^+_2$ state of $^{12}$C (Hoyle state) seems to exhibit the same rainbow pattern 
as that seen in the elastic scattering cross section, and Hamada {\it et al.} 
\cite{Ohk13} suggested that this is due to the extended 3$\alpha$-cluster 
structure of the Hoyle state. 

The weakening or suppression of the Airy oscillation in the inelastic \AA scattering 
remains so far an unsolved problem. As an illustration, we show in Fig.~\ref{f1} the 
elastic and inelastic scattering data measured for the \cc system at $E_{\rm lab}=240$ MeV 
\cite{Boh85,Dem10} and for the \oc system at $E_{\rm lab}=200$ MeV \cite{Ohk14,Ogl00}. 
These elastic scattering data were proven to be strongly refractive, with the first 
Airy minimum A1 followed by a broad rainbow shoulder \cite{Kho16}, especially, 
in the elastic \oc scattering data at 200 MeV shown in the lower panel of Fig.~\ref{f1}. 
While the cross section of the inelastic scattering to the $2^+_1$ state of $^{12}$C 
target is quite strong, A1 disappears in the inelastic cross section and the Airy 
pattern is smeared out. The measured inelastic scattering cross section of the $2^+_1$ 
state is even larger than the elastic scattering cross section at medium and large 
angles, which confirms that the inelastic scattering channel remains strongly 
refractive. Thus, the disappearance of the Airy pattern should not be due to an 
enhanced absorption in the exit $2^+_1$ channel.   

To explore this effect, we propose a compact method for the NF decomposition of
the inelastic scattering amplitude to determine explicitly all partial wave 
contributions to the angular momentum transfer to the spin of the excited target. 
For this purpose, the NF decomposition method by Fuller \cite{Ful75} is extended 
to split the inelastic scattering amplitude of the coupled partial waves into the 
near-side and far-side components, so that the refraction in the inelastic scattering 
channel is studied on equal footing with that in the elastic channel, and the 
formation of the nuclear rainbow therein.  
Given the prominent nuclear rainbow pattern observed in the elastic \cc and \oc 
scattering at the refractive energies around $10\sim 20$ MeV/nucleon, we apply 
the extended NF decomposition method to the inelastic scattering to the $2^+_1$ 
(4.44 MeV) state of $^{12}$C target in the coupled channel (CC) analysis of the inelastic 
scattering data measured for the \cc system at $E_{\rm lab}=240$ MeV \cite{Boh85,Dem10}, 
and the \oc system at $E_{\rm lab}=200$ and 260 MeV \cite{Ohk14,Ogl00}. 

\section{Near-far decomposition of the inelastic scattering amplitude in the CC
 formalism}
\label{sec2} 
\subsection{General formalism}\label{sec2.1}
We recall here briefly the coupled channel equations for the elastic and inelastic 
\AA scattering. The scattering wave function is obtained at each total angular momentum 
$J^\pi$ of the \AA system from the solution of the following CC equations \cite{Kho00} 
\begin{eqnarray}
\left\{\frac{\hbar^2}{2\mu_\beta}\left[\frac{d^2}{dR^2}+k^2-
\frac{L(L+1)}{R^2}\right]-\langle\beta(LI)J |V|\beta(LI)J\rangle \right\}
\ \ \nonumber \\
\times\chi_{\beta J}(k,R)=\sum_{\beta' L'I'}\langle \beta(LI)J|V|\beta'(L'I')J\rangle 
\chi_{\beta' J}(k',R),\ \ \ \label{eq1}
\end{eqnarray}
where $\beta$ and $\beta'$ denote the entrance and exit channels, respectively; 
$\mu_\beta$ is the reduced mass, $\hbar k=\sqrt{2\mu_\beta E_\beta}$ 
is the center-of-mass (c.m.) momentum, and $\chi_{\beta J} (k,R)$ 
is the scattering wave function at the internuclear radius $R$. The total 
angular momentum $J$ is determined from the spin $I$ ($I'$) and orbital momentum 
$L$ ($L'$) of the entrance (exit) channel by the angular momentum addition 
$\bm{J}=\bm{L}+\bm{I}=\bm{L'}+\bm{I'}$. In the present work, we focus on
the scattering of the two spinless nuclei ($I=0$) with a natural-parity excitation 
of the target. Then, $|L'-I'|\leqslant J=L\leqslant L'+I'$, where $I'$ is the spin 
of the excited target.

The diagonal matrix element $V_{\beta\beta}(R)$ of the projectile-target 
interaction in Eq.~(\ref{eq1}) is the \AA OP, and the nondiagonal 
matrix element $V_{\beta\beta'}(R)$ is the transition potential, which is also
dubbed as the inelastic scattering form factor (FF). The OP and inelastic 
scattering FF can be evaluated microscopically in the double-folding model (DFM) 
using the ground-state (g.s.) and transition nuclear densities, respectively, 
and an effective nucleon-nucleon (NN) interaction between the projectile- and 
target nucleons (see more details in Ref.~\cite{Kho00}). From the solution 
of the CC equations (\ref{eq1}), we obtain the elastic scattering amplitude as 
\begin{equation}
 f(\theta)=f_{\rm C}(\theta)+\frac{1}{2ik}\sum_L(2L+1)\exp(2i\sigma_L)
 (S_L-1)P_L\left(\cos\theta\right), \label{eq2}
\end{equation}
where $f_{\rm C}(\theta)$ and $\sigma_L$ are the Rutherford scattering amplitude and 
Coulomb phase shift, respectively; $S_L$ is the diagonal element of the elastic scattering 
$S$ matrix, and $P_L(\cos\theta)\equiv P_{LM=0}(\cos\theta)$ is the Legendre function 
of the first kind. Within the CC formalism \cite{Sat83}, the amplitude of the inelastic 
scattering to an excited state of the target with spin $I'$ and projection $M_{I'}$ 
is written explicitly as 
\begin{eqnarray}
f_{M_{I'}}(\theta,\phi)&=&\frac{\sqrt{4\pi}}{2ik}\sum_{LL'}\sqrt{2L+1}
\langle L'-M_{I'} I' M_{I'}|L 0 \rangle \nonumber \\ 
&& \times \exp[i(\sigma_L+\sigma'_{L'})] S'_{L'L} Y_{L'-M_{I'}}(\theta,\phi). \label{eq3}
\end{eqnarray}
Here $Y_{LM}(\theta,\phi)$ is the spherical harmonics, the Coulomb phase shift 
$\sigma'_{L'}$ is evaluated from the c.m. momentum $k'$ in the exit channel, and 
$S'_{L'L}$ is the element of the inelastic scattering $S$ matrix. The orbital angular 
momenta in the entrance and exit channels are linked with spin $I'$ of the excited 
target by the triangular rule 
\begin{equation}
 L'=L-I',L-I'+2,...,L+I'-2,L+I', \label{eq4}
\end{equation}
where the step of two angular-momentum units is implied by the parity conservation. 

\subsection{Multipole mixing of the partial waves} 
\label{sec2.2}
One can see from the expansion (\ref{eq3}) that the coupled partial waves of different  
multipolarities can contribute coherently to the inelastic scattering amplitude
at the same scattering angle $\theta$ when $I'\neq 0$. By expressing the selection 
rule (\ref{eq4}) as $L'=L+K$, the inelastic scattering amplitude (\ref{eq3}) can be 
written in terms of the $K$-subamplitudes allowed by the selection rule (\ref{eq4}) 
\begin{equation}
 f_{M_{I'}}(\theta,\phi)=\sum_{K=-I'}^{I'} f^{(K)}_{M_{I'}}(\theta,\phi). \label{eq5}
\end{equation} 
Like the elastic amplitude (\ref{eq2}), each $K$-subamplitude of $f_{M_{I'}}(\theta,\phi)$ 
can be expanded over the orbital momenta of the entrance channel $L$ as   
\begin{eqnarray}
 & &f^{(K)}_{M_{I'}}(\theta)=\frac{\sqrt{4\pi}}{2ik}\sum_{L}\sqrt{2L+1}
 \langle (L+K)-M_{I'} I' M_{I'}|L 0 \rangle \nonumber \\ 
 & & \times\exp[i(\sigma_L+\sigma'_{L+K})]S'_{(L+K)L}Y_{(L+K)-M_{I'}}(\theta,\phi). 
 \label{eq6}
\end{eqnarray}
In terms of the inelastic scattering cross section, the contribution from each 
$K$-subamplitude is obtained at the given scattering angle as 
\begin{equation}
 \frac{d\sigma_K}{d\Omega}=\sum_{M_{I'}}
 \left|f^{(K)}_{M_{I'}}(\theta,\phi)\right|^2, \label{eq7}
\end{equation} 
so that the full cross section of the inelastic scattering to an excited 
state of the target with spin $I'$ is 
\begin{equation}
\frac{d\sigma}{d\Omega}=\sum_{M_{I'}}
\left|\sum_{K=-I'}^{I'} f^{(K)}_{M_{I'}}(\theta,\phi)\right|^2.   \label{eq8}
\end{equation} 
Thus, for an excited state with spin $I'\neq 0$, the full inelastic scattering 
cross section is given by the interference of the $K$-subamplitudes 
with $K=-I',I'+2,...,I'-2,I'$.   

\subsection{Near-far decomposition of the scattering amplitude}
As mentioned above, the NF decomposition method by Fuller \cite{Ful75} is a very 
helpful tool to analyze the interference of the near-side and far-side scattering 
amplitudes in the elastic scattering \cite{Bra97,Kho07r}. Namely, the elastic 
scattering amplitude is decomposed into the near-side ($f_{\rm N}$) and far-side 
($f_{\rm F}$) components as
\begin{eqnarray} 
f(\theta)&=&f_{\rm N}(\theta)+f_{\rm F}(\theta)=f^{\rm N}_{\rm C}(\theta)+
 f^{\rm F}_{\rm C}(\theta) \nonumber \\ 
&+&\frac{1}{2ik}\sum_L(2L+1)\exp(2i\sigma_L)(S_L-1)\nonumber \\ 
&\times& \left[\tilde Q_L^{(-)}(\cos\theta)+\tilde Q_L^{(+)}(\cos\theta)\right], 
\label{eq9}
\end{eqnarray} 
where $f^{\rm N(F)}_{\rm C}(\theta)$ is the near-side (far-side) component of the 
Rutherford scattering amplitude \cite{Ful75}, the relative strength of the near-side 
and far-side nuclear scattering is given by $\tilde Q_L^{(-)}(\cos\theta)$ and 
$\tilde Q_L^{(+)}(\cos\theta)$, respectively,  
\begin{equation} 
 \tilde Q_L^{(\mp)}(\cos\theta)=\frac{1}{2} 
 \left[P_L(\cos\theta)\pm {2i\over\pi}Q_L(\cos\theta)\right], \label{eq10}
\end{equation}
where $Q_L(\cos\theta)\equiv Q_{LM=0}(\cos\theta)$ is the Legendre function of the 
second kind. It is well established \cite{Bra97,Kho07r,McV92} that the nuclear 
rainbow pattern observed in the elastic \aA and light HI scattering is determined
entirely by the far-side component of the elastic amplitude (\ref{eq9}). The nuclear
rainbow pattern is a broad oscillation of the Airy minima at medium and large 
scattering angles that results from an interference between $f_{{\rm F}_<}(\theta)$ 
and $f_{{\rm F}_>}(\theta)$ subamplitudes of the far-side component in (\ref{eq9}), 
with $L$ being smaller or larger than a critical partial wave $L_{\rm R}$ associated 
with the rainbow angle $\theta_{\rm R}$ \cite{McV92}.

It is natural to expect a similar rainbow pattern also in the inelastic scattering 
cross section of a \AA system that exhibits a pronounced nuclear rainbow in the 
elastic scattering channel. For this purpose, a NF decomposition of the inelastic 
scattering amplitude (\ref{eq3}) should be done for each projection $M_{I'}$ of the 
target spin $I'$, with the contributions of all allowed $K$-subamplitudes (\ref{eq6})
treated explicitly. However, such a detailed decomposition method is so far 
unavailable, and only some general discussion on possible Airy structure 
of the inelastic scattering cross section was made \cite{Ohk13,Ohk14} based on the 
Airy pattern established in the elastic scattering cross section. We note here an 
early attempt to extend the NF decomposition method to the inelastic HI scattering
by Dean and Rowley \cite{Dean84}, where the near-side and far-side scattering 
amplitudes obtained for each $M_{I'}$ magnetic substate of the target excitation 
with $I'\neq 0$ were shown to be {\it not} in phase. However, the refractive Airy
pattern of the nuclear rainbow was not discussed at all in Ref.~\cite{Dean84}.  
To close this gap of the scattering theory, we suggest in the present work a method 
of the NF decomposition of the inelastic scattering amplitude (\ref{eq3}) to investigate 
explicitly the Airy oscillation pattern in the inelastic scattering cross section. 
Thus, the NF decomposition (\ref{eq9}) is generalized to decompose the inelastic 
scattering amplitude (\ref{eq3}) using the associated Legendre functions as    
\begin{eqnarray} 
& & f_{M_{I'}}(\theta,\phi)=f^{\rm N}_{M_{I'}}(\theta,\phi)+
   f^{\rm F}_{M_{I'}}(\theta,\phi)\nonumber \\ 
&=&\frac{\sqrt{4\pi}}{2ik}\sum_{LL'}\sqrt{2L+1}
\langle L'-M_{I'}I'M_{I'}|L 0 \rangle\nonumber \\ 
&\times& A_{L'M_{I'}}\exp[i(\sigma_L+\sigma'_{L'})]\exp(-iM_{I'}\phi) \nonumber\\
&\times& S'_{L'L}\left[\tilde Q_{L'-M_{I'}}^{(-)}(\cos\theta)+
\tilde Q_{L'-M_{I'}}^{(+)}(\cos\theta)\right]. \label{eq11} \\
{\rm where}& & \tilde Q_{LM}^{(\mp)}(\cos\theta)={1\over 2}
\left[P_{LM}(\cos\theta)\pm {2i\over\pi}Q_{LM}(\cos\theta)\right], \nonumber \\
{\rm and}& & A_{LM}=\sqrt{\frac{2L+1}{4\pi}\frac{(L+M)!}{(L-M)!}}. \nonumber
\end{eqnarray}   
Here $P_{LM}(\cos\theta)$ and $Q_{LM}(\cos\theta)$ are the associated Legendre 
functions of the first- and second kind, respectively. We note that the 
inelastic scattering FF includes both the Coulomb and nuclear contributions 
\cite{Kho00}, and the inelastic Coulomb scattering amplitude is not treated 
separately as in the elastic scattering channel. 

Expressing $L'=L+K$ in the generalized NF decomposition (\ref{eq11}), we obtain 
explicitly the near-side and far-side components of each $K$-subamplitude of the 
inelastic scattering amplitude (\ref{eq5}) as
\begin{eqnarray} 
& & f^{(K)}_{M_{I'}}(\theta,\phi)=f^{(K)}_{{\rm N}, M_{I'}}(\theta,\phi)+
   f^{(K)}_{{\rm F}, M_{I'}}(\theta,\phi)\nonumber \\ 
&=&\frac{\sqrt{4\pi}}{2ik}\sum_L\sqrt{2L+1}
\langle (L+K)-M_{I'}I'M_{I'}|L 0\rangle\nonumber \\ 
&\times& A_{(L+K)M_{I'}}\exp[i(\sigma_L+\sigma'_{L+K}-M_{I'}\phi)] S'_{(L+K)L}\nonumber \\ 
&\times& \left[\tilde Q_{(L+K)-M_{I'}}^{(-)}(\cos\theta)+
\tilde Q_{(L+K)-M_{I'}}^{(+)}(\cos\theta)\right]. \label{eq12}
\end{eqnarray}   
Thus, the generalized NF decomposition (\ref{eq12}) allows us to determine the near-side 
and far-side contributions from each $K$-subamplitude to the partial (\ref{eq7}) and 
full inelastic cross section (\ref{eq8}), and to study the formation of the nuclear 
rainbow in the inelastic \AA scattering in the same manner as done for the elastic 
scattering using the Fuller method (\ref{eq9}). 

\section{Elastic and inelastic \cc and \oc scattering at the refractive energies}
 
The diagonal matrix element $V_{\beta\beta}(R)$ of the projectile-target interaction 
in the CC equations (\ref{eq1}) is determined by the total optical potential $U_0(R)$
\cite{Kho00}. The new version of the density dependent CDM3Y3 interaction with the 
rearrangement term included \cite{Kho16} is used in the double-folding calculation 
of the real optical potential $V_0(R)$. Because the nuclear rainbow is a subtle 
effect that can be observed only when the absorption of the dinuclear system is 
weak, the imaginary OP in the flexible Woods-Saxon (WS) form is usually used for
a proper identification of the rainbow pattern. Thus, we have  
\begin{eqnarray}
 &U_0(R)&=N_RV_0(R)+ iW_0(R)+V_{\rm C}(R),\nonumber \\ 
\ \ \ {\rm where}\ \ &W_0(R)&=-\frac{W_V }{1+\exp[(R-R_V)/a_V]}. \label{eq14} 
\end{eqnarray}
The Coulomb potential $V_{\rm C}(R)$ is obtained by folding the two uniform 
charge distributions with their mean-squared radii chosen to be close to the measured 
charge radii of the two nuclei. The nuclear g.s. densities used in the DFM calculation 
are taken as the Fermi distributions with parameters chosen to reproduce the empirical 
matter radii of the considered nuclei \cite{Kho01}. The renormalization $N_R$ of the 
real folded potential and the WS parameters (\ref{eq14}) are adjusted in each case 
to the best CC description of the elastic scattering data, and a small deviation 
of $N_R$ from unity validates the use of the folding model. The best-fit OP parameters 
used in the present CC study of the elastic and inelastic 
\cc and \oc scattering are given in Table~\ref{t1}. 
\begin{table*}[bht]
\begin{center}
\caption{The OP parameters (\ref{eq14}) used in the CC calculation (\ref{eq1}) of the 
elastic and inelastic  \cc and \oc scattering. $J_R$ and $J_W$ are the volume 
integrals (per interacting nucleon pair) of the real and imaginary parts of the OP, 
respectively.} \label{t1} 
\vspace*{0cm}\hspace*{0cm}
		\begin{tabular}{ccccccccc} 
		\hline\hline\
			& $E_{\rm lab}$ & $N_R$ & $J_R$ & $W_V$ & $R_V$ & $a_V$ & $J_W$ & Data \\
			& (MeV) &  & (MeV~fm$^3$) & (MeV) & (fm) & (fm) & (MeV~fm$^3$) & \\ \hline
			\cc	&240 & 1.067 & 336.0 & 19.29  &	5.743 &	0.595 & 117.5	& \cite{Boh85,Dem10} \\ 
			\oc	&200 & 0.936 & 300.4 & 13.32  & 6.150 &	0.502 &	72.06 & \cite{Ohk14,Ogl00} \\ 	
			\oc &260 & 0.930 & 291.8 & 18.50  &	5.756 &	0.550 &	83.92 & \cite{Ohk14,Ogl00} \\		
			\hline\hline 
		\end{tabular}
\end{center}
\end{table*} 

The nondiagonal matrix element $V_{\beta\beta'}(R)$ in the CC equations (\ref{eq1}) 
is given by the inelastic form factor $U_{I'}(R)$ that accounts for inelastic 
scattering to the target excited state with spin $I'$ (see details in Ref.~\cite{Kho00}), 
\begin{equation}
  U_{I'}(R)=N_RV_{I'}(R)-i\delta_{I'}\frac{\partial W_0(R)}
 {\partial R}+ V_{{\rm C},I'}(R), \label{eq15} 
\end{equation}
where the real nuclear $V_{I'}(R)$ and Coulomb $V_{{\rm C},I'}(R)$ inelastic form
factors are calculated in the DFM using the nuclear transition densities of the 
excited states of $^{12}$C obtained in the resonating group method (RGM) \cite{Kam81}. 
The nuclear deformation lengths $\delta_{I'}$ are determined by the collective-model 
prescription using the measured $B(EI')$ transition rates \cite{Ram87,Kib02} of 
the considered excited states of $^{12}$C. All the CC calculations have been  
done using the code ECIS97 written by Jacques Raynal \cite{Ray72} that provides
the detailed output of the elastic and inelastic scattering matrices necessary for
the NF decomposition (\ref{eq9}) and (\ref{eq12}) of the corresponding scattering 
amplitudes. It should be recalled that the sequential iteration method implemented in 
the ECIS code was developed by Raynal to tackle the inelastic HI scattering with 
a strong Coulomb contribution, focusing in particular on the scattering experiments 
being carried out at GANIL at that time \cite{Alamanos20}. By using the recurrence 
relations for the Coulomb excitation integrals in the CC calculations \cite{Ray81}, 
the ECIS integration of the coupled equations is highly stable and accurate up to 
very large radii with sufficiently high number of partial waves. For the \cc and \oc 
systems under the present study, the ECIS integration up to $R_{\rm max}\approx 25$ fm 
in steps of $dR=0.05$ fm is needed to ensure the convergence of the calculated cross 
section, taking into account up to 180 partial waves. At the considered energies, 
the CC results obtained using the nonrelativistic and relativistic kinematics are
about the same.  

The elastic and inelastic \cc scattering has been widely studied at energies ranging
from the Coulomb barrier \cite{Row77} up to about 200 MeV/nucleon \cite{Hos88}. While
the elastic \cc scattering at the barrier energies was shown to be of interest for  
nuclear astrophysics \cite{Chie18}, the scattering data measured for this system at
the refractive energies around 20 MeV/nucleon \cite{Boh85,Dem10} exhibit a nuclear 
rainbow pattern that enabled an unambiguous determination of the real OP down to 
small distances \cite{Kho07r,Kho16}. In particular, the \cc scattering data measured 
accurately at $E_{\rm lab}=240$ MeV \cite{Boh85,Dem10} are very important for our 
study because this energy was found optimal for the observation of the first Airy 
minimum A1 of the nuclear rainbow in the elastic \cc scattering \cite{Kho16}. 
As shown above in Fig.~\ref{f1}, the data of the inelastic \cc scattering to the 
$2^+_1$ (4.44 MeV) state of $^{12}$C measured at this same energy \cite{Boh85,Dem10} 
does not have any minimum that can be interpreted as the remnant of A1, at angles  
near the location of A1 established in the elastic cross section. Another light HI 
system that exhibits a prominent rainbow pattern in the elastic scattering is \oc
\cite{Ogl00}. Unlike $^{12}$C+$^{12}$C, the \oc system does not have the boson symmetry, 
and the angular evolution of the Airy pattern could be observed with the increasing 
energy. The strongest rainbow pattern, the deep A1 minimum followed by an exponential 
fall-off of the rainbow shoulder, is well confirmed in the elastic \oc scattering
data measured at $E_{\rm lab}=200$ MeV \cite{Ogl00}. The question why the inelastic 
\oc scattering data measured at this same energy \cite{Ohk14} does not show a 
similar Airy pattern (see lower panel of Fig.~\ref{f1}) is so far unanswered. 
In the present work, we try to explain the suppression of the first Airy minimum 
in the inelastic \cc and \oc scattering cross sections at the energies where A1 
was clearly identified in the elastic cross sections measured for these systems. 

\begin{figure}[bht]\hspace*{-0.2cm}
\includegraphics[width=0.5\textwidth]{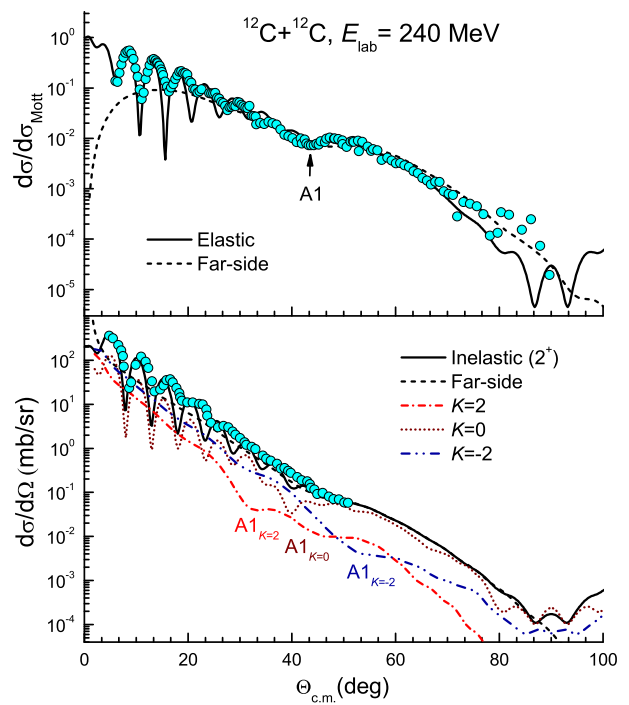}\vspace*{0cm}
\caption{CC description (solid lines) of the elastic and inelastic (to the
$2^+_1$ state of $^{12}$C) \cc scattering at $E_{\rm lab}=240$ MeV, in comparison 
with the measured data \cite{Boh85,Dem10}. The dash-dotted, dotted, and 
dash-dot-dotted lines are the partial inelastic cross sections (\ref{eq7}) 
given by the subamplitudes with $K=2,0$ and -2, respectively. The dashed lines 
are the far-side cross sections given by the NF decompositions (\ref{eq9}) and 
(\ref{eq11}) of the elastic and inelastic scattering amplitudes. The first 
Airy minimum A1 is shown explicitly for each partial inelastic 
cross section.}  \label{f2}
\end{figure}
The CC results for the elastic and inelastic \cc scattering describe well the 
data as shown in Fig.~\ref{f2}. The dominance of the far-side cross sections 
at medium and large angles indicates that both the elastic- and inelastic \cc 
at the considered energies are strongly refractive. The nuclear rainbow pattern 
is well established in the elastic \cc scattering, with the first Airy minimum A1 
unambiguously identified \cite{Kho16} at the scattering angle 
$\theta_{\rm c.m.}\approx 41^\circ$ based on the NF decomposition (\ref{eq9}) 
of the elastic scattering amplitude. The data of the inelastic \cc scattering 
to the $2^+_1$ state of $^{12}$C are reproduced reasonably by the CC calculation 
(\ref{eq1}), but the Airy structure seen in the elastic cross section is  
smeared out in the inelastic cross section. Given the dominance of the far-side 
scattering at medium and large angles, the suppression of A1 in the inelastic 
cross section is definitely \emph{not} caused by the near-side/far-side interference, 
but more likely by a destructive interference of the far-side subamplitudes.  
\begin{figure}[bht]\hspace*{-0.2cm}
	\includegraphics[width=0.5\textwidth]{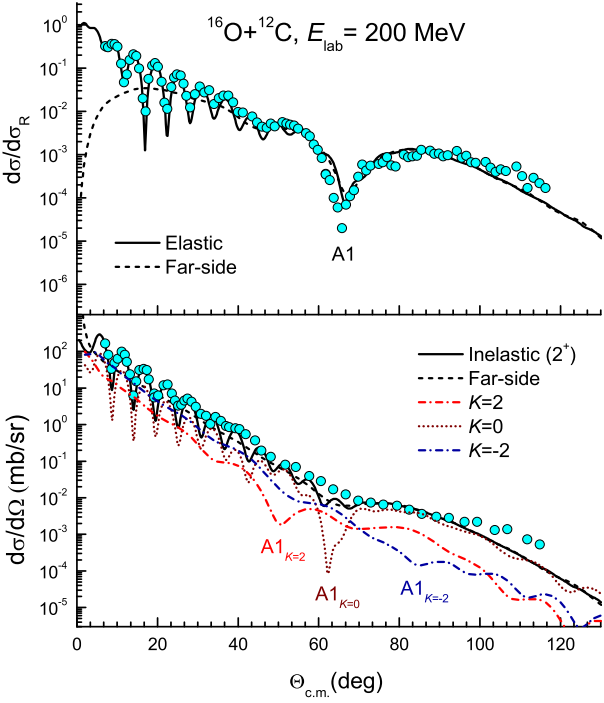}\vspace*{-0.0cm}
	\caption{The same as Fig.~\ref{f2} but for the elastic and inelastic \oc 
	scattering at $E_{\rm lab}=200$ MeV \cite{Ohk14,Ogl00}.} \label{f3}
\end{figure}  
\begin{figure}[bht]\hspace*{-0.2cm}
	\includegraphics[width=0.5\textwidth]{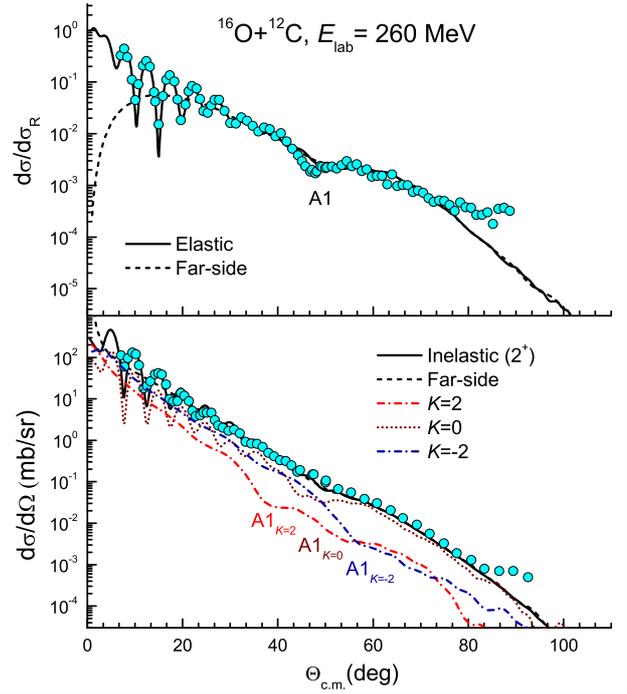}\vspace*{-0.0cm}
	\caption{The same as Fig.~\ref{f2} but for the elastic and inelastic \oc 
	scattering at $E_{\rm lab}=260$ MeV \cite{Ohk14,Ogl00}.} \label{f4}
\end{figure}

One can see in the partial-wave expansion of the inelastic scattering amplitude 
(\ref{eq3}) that the inelastic scattering cross section at the given scattering 
angle contains the contributions from the subamplitudes of different partial  
waves ($L'\neq L$) when the spin of the excited state is nonzero ($I'\neq 0$). For the 
inelastic \cc scattering to the $2^+_1$ state of $^{12}$C shown in Fig.~\ref{f2}, 
each $L$-component of the inelastic scattering amplitude is resulting from an 
interference  of the three $K$-subamplitudes with $K=L'-L=-2,0,2$. The partial 
inelastic scattering cross sections (\ref{eq7}) given by the three $K$-subamplitudes 
(summed over all partial waves $L$) are shown separately in the lower panel of 
Fig.~\ref{f2}. By tracing the angular evolution of the corresponding far-side 
cross sections, we have identified the first Airy minimum A1 in the partial 
inelastic \cc cross section with $K=0$ at $\theta_{\rm c.m.}\approx 40^\circ$ 
which is close to the locations of A1 in the elastic \cc cross section.
While a slight remnant of A1 with $K=0$ can still be seen in the calculated 
inelastic scattering cross section (\ref{eq8}), it is not observed in the measured 
data. Rather weak rainbow patterns of the two partial inelastic \cc cross sections 
with $K\neq 0$ were found which are shifted in angles, with A1 located at 
$\theta_{\rm c.m.}\approx 33^\circ$ and $52^\circ$ in the partial inelastic 
cross section with $K=2$ and $K=-2$, respectively. One can see in the lower panel
of Fig.~\ref{f2} that the partial inelastic cross sections with $K=\pm 2$ 
are much weaker than that with $K=0$ at medium and large angles, and no remnants 
of A1 with $K=\pm 2$ can be seen in the total inelastic cross section.

A similar picture can be seen in the CC results for the elastic and inelastic 
\oc scattering at $E_{\rm lab}=200$ MeV plotted in Fig.~\ref{f3}. While the 
prominent A1 minimum is located at $\theta_{\rm c.m.}\approx 65^\circ$ in the 
elastic cross section, it seems to disappear in the inelastic \oc scattering 
cross section. Such an effect was found also in the results of the earlier CC 
analysis of the inelastic \oc scattering \cite{Ohk17} as well as those of a cluster 
folded model study \cite{Has18}. As discussed for the \cc system, three different
Airy oscillation patterns can be seen in the partial inelastic \oc cross sections 
given by the subamplitudes with $K=2,0,-2$, with A1 located at 
$\theta_{\rm c.m.}\approx 49^\circ, 62^\circ$, and $84^\circ$, respectively. 
Again, the location of A1 with $K=0$ is quite close to the locations of A1 
in the elastic \oc cross section. The full inelastic scattering cross section 
(\ref{eq8}) includes the contributions from all allowed $K$-subamplitudes, and 
their out-of-phase interference smears out the individual A1 minima seen in the 
partial inelastic cross sections (\ref{eq7}). While a slight remnant of A1 with 
$K=0$ is seen in the calculated inelastic cross section (solid line in the lower 
panel of Fig.~\ref{f3}), it cannot be clearly resolved in the measured data. 
The same CC results for the elastic and inelastic \oc scattering at $E_{\rm lab}=260$ 
MeV are compared with the data \cite{Ohk14,Ogl00} in Fig.~\ref{f4}. We found that 
the absorption becomes slightly stronger (see Table~\ref{t1}) with the increasing 
energy, and the Airy oscillation pattern is weakened and shifted to smaller angles 
(with A1 in the elastic cross section located at $\theta_{\rm c.m.}\approx 50^\circ$). 
The weaker Airy oscillation pattern of each partial inelastic cross section can 
still be seen but the remnant of A1 with $K=0$ disappears in both the calculated 
inelastic scattering cross section and measured data (lower panel of Fig.~\ref{f4}). 
In conclusion, the results of our CC analysis shown in Figs.~\ref{f2}-\ref{f4} 
explain naturally why the Airy oscillation pattern of the nuclear rainbow is 
strongly suppressed in the measured data of the inelastic \cc and \oc scattering 
to the $2^+_1$ state of $^{12}$C at the rainbow energies.   

\begin{figure}[bht]\hspace*{-0.2cm}
\includegraphics[width=0.5\textwidth]{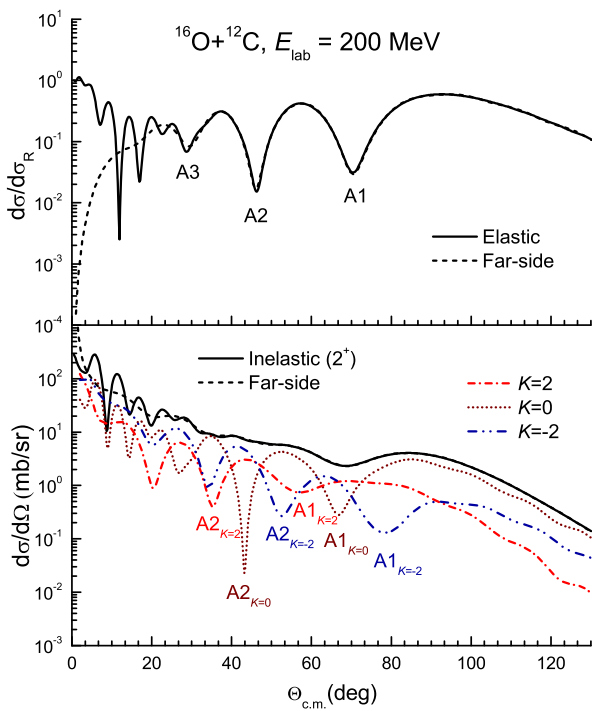}\vspace*{0.0cm}
\caption{The same CC results as those in Fig.~\ref{f3} for the elastic and 
inelastic \oc scattering at $E_{\rm lab}=200$ MeV, obtained with a less absorptive 
OP (\ref{eq14}) with $W_V\to W_V/3$.} \label{f5}
\end{figure}
It is well-known that the nuclear rainbow is formed by the interference 
of the far-side scattering waves, refracted by the attractive \emph{real} OP 
\cite{Bra97,Kho07r}. That's the reason why the nuclear rainbow could be observed
only when the absorption of the scattering system is weak enough for the far-side
trajectories to survive at the medium and large scattering angles. In practice, the 
absorptive strength of the OP is often reduced to artificially enhance the far-side 
scattering amplitude for a proper identification of the Airy oscillation pattern 
\cite{Kho16}. The results of the CC calculation of the elastic and inelastic \oc 
scattering at 200 MeV given by a less absorptive OP (with $W_V\to W_V/3$) are plotted 
in Fig.~\ref{f5}, and one can trace in the elastic cross section the whole pattern 
of the nuclear rainbow including the first (A1), second (A2), and third (A3) Airy 
minima. The same Airy oscillation pattern can be seen also in the three partial 
inelastic scattering cross sections (\ref{eq7}) given by the three $K$-subamplitudes 
(\ref{eq6}), but the locations of the Airy minima are shifted to the smaller angles 
when $K=2$, and to the larger angles when $K=-2$. It is very essential to emphasize 
again that the Airy oscillation pattern in the partial inelastic cross section with 
$K=0$ remains about the same as that observed in the elastic scattering thanks to 
an in-phase interference of the partial waves with $L'=L$. When $K\neq 0$, the 
out-of-phase interference of the partial waves with $L'\neq L$ smears out the 
different Airy oscillation patterns in the full inelastic $2^+_1$ scattering 
cross section. Because the partial inelastic cross section with $K=0$ is substantially 
larger that those with $K=\pm 2$ at medium and large angles, the remnant of the 
first Airy minimum A1 with $K=0$ can be very well seen in the full inelastic 
scattering cross section (solid line in the lower panel of Fig.~\ref{f5}) when 
a reduced absorption $W_V$ was used in the CC calculation. In fact, the broad 
rainbow shoulder following A1 with $K=0$ is still visible in the data measured at 
$E_{\rm lab}=200$ MeV for the inelastic \oc scattering to the $2^+_1$ state 
of $^{12}$C (see lower panel of Fig.~\ref{f3}). 

It becomes clear now that there is no unique Airy pattern of the nuclear rainbow 
in the full (far-side) cross section of the inelastic \AA scattering to an excited 
nuclear state with nonzero spin. In such a case, only the Airy oscillation pattern 
of the partial inelastic cross section (\ref{eq7}) given separately by each 
$K$-subamplitude (\ref{eq6}) can be determined in the same manner as done in the 
case of elastic scattering. Thus, the detailed locations of the Airy minima A1, A2, 
and A3 in the inelastic \oc scattering cross section deduced visually by Ohkubo 
{\it et al.} from the calculated angular distribution \cite{Ohk17} are not 
properly founded. 
\begin{figure}[bht]\hspace*{0cm}
\includegraphics[width=0.5\textwidth]{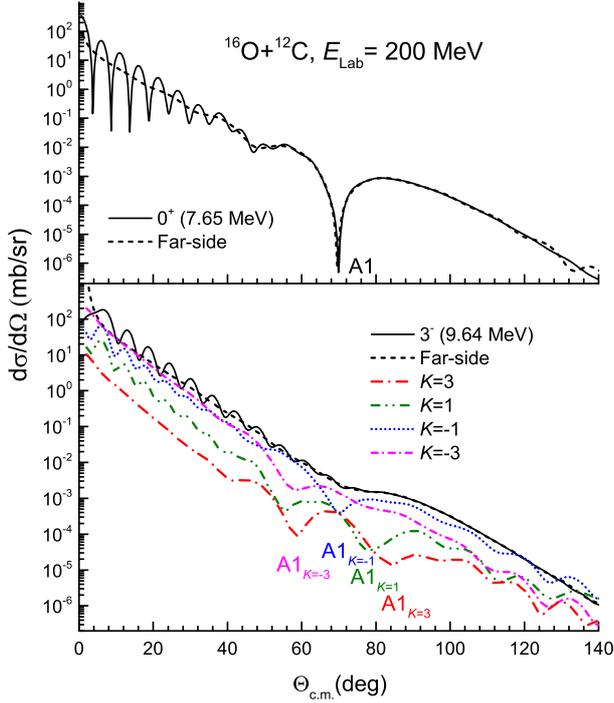}\vspace*{0cm}
\caption{The same CC results as those in Fig.~\ref{f2} but for the inelastic 
\oc scattering to the $0^+_2$ (upper part) and $3^-_1$ (lower part) excited 
states of $^{16}$O at $E_{\rm lab}=200$ MeV.} \label{f6}
\end{figure}

Although the inelastic \oc scattering to the $0^+_2$ (Hoyle) and $3^-_1$ excited
states (at $E_x\approx 7.65$ and 9.64 MeV, respectively) were not measured at 
$E_{\rm lab}=200$ MeV, the CC prediction of the inelastic cross sections for 
these states should be of interest for the revelation of the nuclear rainbow 
pattern therein. The CC results for the inelastic \oc scattering at 200 MeV 
obtained with the nuclear transition densities of the $0^+_2$ and $3^-_1$ states 
of $^{12}$C given by the RGM \cite{Kam81} and the same OP as given in Table~\ref{t1} 
are shown in Fig.~\ref{f6}. One can see that the refractive (far-side) scattering 
is also dominant at medium and large angles in the inelastic scattering to the 
Hoyle ($0^+_2$) state, with $K\equiv 0$ (or $L'\equiv L$) and 
$d\sigma_K/d\Omega\equiv d\sigma/d\Omega$. In this case, spin of the excited
state is zero and there is no interference of the scattering subamplitudes with 
different $K$. As a result, the Airy pattern in the angular distribution of the 
inelastic scattering to the Hoyle state is determined with a single ($K=0$) 
inelastic scattering amplitude, in the same manner as done for the elastic \oc 
scattering. Thus, the deep minimum of the inelastic $0^+_2$ cross section can be 
confirmed as the first Airy minimum A1 which is located at about the same angle as 
A1 of the elastic cross section (see upper panels of Figs.~\ref{f5} and \ref{f6}). 

For the inelastic \oc scattering to the $3^-_1$ state of $^{12}$C, there are four 
subamplitudes (\ref{eq6}) with $K=-3\ (L'=L-3), K=-1\ (L'=L-1),\ K=1\ (L'=L+1)$, and 
$K=3\ (L'=L+3)$ with the corresponding A1 minima shifted away from each other by 
a few degrees in the scattering angle. Among these $K$-subamplitudes, The first Airy 
minimum A1 of the partial inelastic cross section with $K=-1$ is quite pronounced. 
One can see in the lower panel of Fig.~\ref{f6} that the out-of-phace interference 
of these 4 subamplitudes results in a rather smooth full inelastic scattering 
cross section (\ref{eq8}). This is the same effect of suppression of the Airy pattern 
as discussed above for the inelastic \oc scattering to the $2^+_1$ state of $^{12}$C. 

\section{Summary}
The present work explains why the Airy pattern of nuclear rainbow is suppressed 
in the inelastic \cc and \oc scattering to the $2^+_1$ state of $^{12}$C 
at the refractive energies, where a strong rainbow pattern has been observed 
in the elastic scattering. For this purpose, the near-far decomposition 
method by Fuller is generalized to determine the near-side and far-side components 
of the inelastic scattering amplitude for all partial wave contributions. 
Using the generalized NF decomposition method, our coupled channel analysis 
of the elastic and inelastic \cc and \oc scattering at the energies under study 
shows unambiguously that the destructive interference of the inelastic partial 
waves of different multipoles suppresses the Airy oscillation pattern 
in the inelastic scattering cross section. Nevertheless, the inelastic scattering 
remains strongly refractive in these cases, with the dominant far-side scattering 
at medium and large scattering angles. 

We conclude, therefore, that it is not possible to identify uniquely the Airy pattern 
of the nuclear rainbow in the angular distribution of the inelastic \AA scattering 
to an excited state with nonzero spin. Semiclassically, such a refractive mixing 
of the partial waves of different multipoles is analogous to an optical prism refracting 
ray of light of different wave lengths. The only exception is the inelastic scattering to 
a monopole excitation which does not mix different multipoles in the inelastic scattering 
amplitude, and the Airy pattern in the inelastic cross section can be determined 
consistently in the same manner as done for the elastic scattering. In light of this 
result, an accurate measurement of the inelastic $\alpha$ or light ion scattering 
to the $0^+_2$ excitation of the $^{12}$C target should be of interest for the future
studies of the nuclear rainbow scattering as well as the $\alpha$-cluster 
structure of the Hoyle state \cite{Kho08,Ohk04}. 

Last but not least, we gratefully notice that over the years our nuclear scattering 
study has been relied on several versions of the coupled channel code ECIS written by
Jacques Raynal. This state-of-the-art computer code of nuclear scattering is still being 
actively used in the community, and Jacques' important contribution to the development
of the nuclear physics research is strongly appreciated by many of us. 

\section*{Acknowledgments}
The present research was supported, in part, by the National Foundation for Science 
and Technology Development (NAFOSTED). The authors also thank Prof. M. Kamimura for 
his permission to use the RGM nuclear densities in the double-folding calculation.

\end{document}